\documentclass[reprint, aps, nofootinbib]{revtex4-2}
\usepackage{bm}
\usepackage[T1]{fontenc}
\usepackage[utf8]{inputenc}
\usepackage{amsmath, amsthm, amsfonts, amssymb}
\DeclareMathOperator{\arcsinh}{arcsinh}
\usepackage{hyperref}
\usepackage{multirow}
\usepackage{graphicx}

\begin{document}

\title{Adiabatic approach to the trans-Planckian problem in Loop Quantum Cosmology}
 
\author{Luis J. Garay}
\email{luisj.garay@ucm.es}
\author{Mario López González}
\email{mariol22@ucm.es}
\author{Mercedes Martín-Benito}
\email{m.martin.benito@ucm.es}
\affiliation{Departamento de F\'isica Te\'orica and IPARCOS, Facultad de Ciencias F\'isicas, 
Universidad Complutense de Madrid,  Plaza de las Ciencias 1, 28040 Madrid, Spain}  
\author{Rita B. Neves}
\email{rita.neves@sheffield.ac.uk}
\affiliation{Departamento de F\'isica Te\'orica and IPARCOS, Facultad de Ciencias F\'isicas, 
Universidad Complutense de Madrid,  Plaza de las Ciencias 1, 28040 Madrid, Spain}
\affiliation{School of Mathematics and Statistics, University of Sheffield, Hicks Building, Hounsfield Road, Sheffield S3 7RH, United Kingdom}

\begin{abstract}
{We study the scalar modes that, being observable today, were trans-Planckian before inflation, within the context of hybrid Loop Quantum Cosmology (LQC). We analyse the dynamics of these highly ultraviolet modes by introducing modified dispersion relations to their equations of motion and  discuss the impact that these relations would introduce in the power spectrum by  computing the adiabaticity coefficient. More precisely, we consider two different models compatible with observations for the standard linear dispersion relation which are based on different initial conditions for the perturbations and background. One of these models avoids the issue altogether by generating less $e$-folds of inflation, so that the observable modes are never trans-Planckian, whereas the other  suffers (arguably softly) from the trans-Planckian problem. This shows that the existence of the trans-Planckian problem in LQC is model-dependent.

}

\end{abstract}

~\hfill IPARCOS-UCM-24-011

\maketitle
\flushbottom

\section{Introduction}\label{Sec:Introduction}

The classical cosmological description of the evolution of the Universe in terms of the old big bang model starts to fail as one goes back in time. This is reflected in the existence of different problems that arise in such a framework, namely, the flatness, horizon, and monopole problems~\cite{Weinberg:2008zzc}. The solution to them comes from the introduction of an early stage of the Universe with accelerated expansion, known as inflation~\cite{Guth:1980zm, Linde:1981mu}. In fact, inflationary theories are not only good models because they can solve these problems. The main success of inflation is that it describes and explains the origin of the cosmological perturbations, measured in the Cosmic Microwave Background (CMB), through quantum fluctuations of the scalar and gravitational fields. The amplitude of these perturbations is determined by the power spectrum of primordial fluctuations, which turns out to be nearly scale invariant~\cite{Mukhanov:1990me}.

Although the inflationary theory succeeds in explaining the issues mentioned above, it also presents different problems. One of them, which will be the focus of this work, is the so-called trans-Planckian problem of inflationary cosmology~\cite{BranMart_TransPlanckianProblem}. In most current models, inflation involves a huge expansion of the Universe in order to solve the classical problems, which means that physical wavelengths that correspond to large-scale structures we observe today in the CMB were much smaller than the Planck length at the onset of inflation. This questions the validity of the standard results concerning the power spectrum and forces one to consider trans-Planckian effects.

This problem is analogous to the trans-Planckian problem of black hole Physics~\cite{Jacobson:1999zk}. In this context, it was shown that the thermal Hawking spectrum of black holes is robust against modifications of Physics in the high-energy sector, which were represented by modified dispersion relations that deviate from the standard one above some ultraviolet scale~\cite{Unruh:1994je, Corley:1996ar}. 

In inflationary cosmology, the problem has been analysed following the same approach~\cite{BranMart_TransPlanckianProblem, BranMart_Robustness, BranMart_Adiabaticity, BranMart2002, Starobinsky:2001kn,KempfNiem, NiemPar_Adiabaticity, ParMacher_SlowRoll}. The equations of motion of perturbations were changed by introducing modified dispersion relations. The robustness of the predictions of classical cosmology was then analysed by means of the so-called adiabaticity coefficient~\cite{BranMart_Adiabaticity, NiemPar_Adiabaticity}, which accounts for the adiabatic evolution of perturbations modes while they are trans-Planckian. The result is that as long as there is scale separation between the ultraviolet scale and the rate of expansion of the Universe, the imprint in the power spectrum will be negligible, provided that the modified dispersion relation is monotonic.

However, all these analyses ignore the pre-inflationary evolution of the Universe. Naturally, if trans-Planckian Physics is important in inflationary theories, so it will in a pre-inflationary stage, since the Universe is smaller and hence physical wavelengths too. There are different theories that attempt to introduce a pre-inflationary description of the Universe. Among them, Loop Quantum Cosmology (LQC)~\cite{AshtekarLQC_Review} has become in  recent years a promising approach, given its predictive power and the capacity to compute power spectra within this framework~\cite{Agullo_SummaryPertLQC}. The main result of LQC is that it removes the initial singularity of classical cosmology by means of a quantum bounce, due to quantum geometry effects. This leads to  effective modified Einstein equations for the background and provides a well-defined evolution. With this description of the background, one can introduce cosmological perturbations and compute predictions for the primordial power spectra that can be contrasted with observations. In fact, several proposals have been found to be compatible with current observations \cite{Cosmictango,MRJ}. So far, cosmological perturbations have been introduced in this framework by means of a Fock quantisation, as in standard cosmology, while the techniques of LQC are only considered for the quantisation of the background. This leads to an effective description of the propagation of perturbations, where trans-Planckian effects originating from the quantum nature of geometry are only obtained for the background. Although these influence the propagation of the perturbations, they are not the corrections to trans-Planckian modes one expects to obtain from a full theory of quantum gravity. Thus, just as in standard cosmology, one may ask whether these scenarios suffer from the trans-Planckian problem.

In this work, we study the trans-Planckian problem considering pre-inflationary scenarios described by LQC, following the same steps that have been taken in the context of inflation. In particular, we consider two different background evolutions introduced in~\cite{AshtekarValues} and~\cite{MRJ}. Both models are compatible with current observations for the standard linear dispersion relation (see~\cite{Cosmictango} for the analysis regarding the first model). This will allow us to determine whether the predictions of the primordial power spectrum are robust when these scenarios are taken into consideration.

We find that the first model suffers from the trans-Planckian problem, as the adiabaticity coefficient of some of the observable modes becomes non-negligible while they are trans-Planckian prior to inflation. This means that the observable portion of the primordial power spectrum is somewhat sensitive to trans-Planckian Physics in this model. In contrast, the second model circumvents the problem altogether by generating less $e$-folds of inflation, so that the observable modes are never trans-Planckian in the early Universe. This suggests that the existence of the trans-Planckian problem in LQC is model-dependent. This work also motivates the future study of the trans-Planckian problem from deep, fully-geometrical arguments, rather than by introducing modifications to the theory by hand, in order to delve into this subject from a theoretical point of view.

The remainder of this work is structured as follows. In Sec.~\ref{Inflation} we briefly review the inflationary theory and the origin of primordial fluctuations that can be measured in the CMB, whose amplitude is fixed by the power spectrum. In Sec.~\ref{TP Inflation} we give an overview of the trans-Planckian problem of inflationary cosmology and the methodology used to study it. In Sec.~\ref{LQC} we introduce the dynamics of LQC both for the background and the perturbations, and proceed to analyse the trans-Planckian problem within LQC. Sec.~\ref{Numerical Results} is devoted entirely to the numerical results and their discussion for the particular LQC models under consideration. Finally, in Sec.~\ref{Conclusions} we summarize the main results of this work, its limitations and future research paths in this area. The convention used in this work is $\hbar = c = 1$ and $m_\textup{Pl}^{2} = 1/G$. We also use Planckian units: $\ell_{\textup{Pl}} = t_{\textup{Pl}} = m_{\textup{Pl}}^{-1}$.

\section{Inflation and generation of fluctuations}\label{Inflation}

\subsection{Inflation and slow-roll regime}

The main idea of inflation~\cite{Guth:1980zm,Linde:1981mu} is that the scale factor evolves nearly exponentially in cosmological time, and the Universe is suffering an accelerated expansion. This can be achieved in many ways, but the usual and {{simpler}} approach is based on the existence of a scalar field, known as inflaton~\cite{Belinsky:1985zd}. An inflaton $\phi(t)$ can be described as a perfect fluid, so that its energy density and pressure are:
\begin{equation}\label{energydensity&pressure}
    \rho=\dot{\phi}^2 / 2 + V(\phi), \qquad p=\dot{\phi}^2 / 2 - V(\phi),
\end{equation}
where $V(\phi)$ is the potential of the inflaton and the dot means derivative with respect to cosmological time $t$. Hence, if one assumes that the Universe is filled with an inflaton, two independent equations of motion arise for a Friedmann-Lemaître-Robertson-Walker (FLRW) Universe, namely:
\begin{align}
H^{2} = \frac{8\pi }{3 m_{\textup{Pl}}^2} \rho , \qquad
        \ddot{\phi} + 3H(t)\dot{\phi} + V'(\phi) = 0,
    \label{backgroundeqns}
\end{align}
where $H(t)=\dot{a}(t)/a(t)$ is the Hubble parameter and $a(t)$ is the scale factor. In order to get inflation one must impose $p \approx -\rho$, that is to say, that the condition $\dot{\phi}^{2} \ll V\left(\phi\right)$ for the inflaton holds during a sufficiently long period of time. This can be achieved if the term $\ddot{\phi}$ is negligible, which is commonly referred to as the slow-roll regime. When this slow-roll regime ceases to be valid inflation ends, the inflaton begins to oscillate, and the Universe starts decelerating. 

The so-called number of $e$-folds $N=\log[a(t_{f})/a(t_{i})]$ quantifies the exponential increase, where $t_i$ and $t_f$ are the times when inflation starts and ends.

\subsection{Primordial fluctuations}\label{Inflation B}
We consider linear cosmological perturbations around a homogeneous background (see~\cite{Mukhanov:1990me} for an exhaustive review) which explain the origin of primordial fluctuations measured in the CMB by means of the power spectrum.
We will focus on scalar perturbations, as they are the ones that leave that observable imprint in the CMB. Tensor perturbations admit a similar analysis but are related to primordial gravitational waves, which have not been observed yet, while vector perturbations are diluted in cosmological evolution~\cite{Mukhanov:2005sc}.

Due to the gauge freedom and the connection between metric and matter perturbations through the Einstein equations, scalar perturbations can be described by means of a single degree of freedom that we will take as the Mukhanov-Sasaki variable $v=z\mathcal{R}$. Here $\mathcal{R}$ is the comoving curvature perturbation, which is gauge invariant and accounts for metric and matter perturbations, and $z= a\dot{\phi} / H$.
Up to linear order in perturbations and decomposing $v$ in Fourier modes, one gets the Mukhanov-Sasaki equation~\cite{Mukhanov:1990me}:
\begin{equation}\label{muksaseqn}
    v_{k}'' + \omega_{k}^{2}(\eta) v_{k} = 0, \qquad \omega_{k}^{2}(\eta) = k^{2} - z'' / z ,
\end{equation}
where $k=|\vec{k}|$ is the comoving wavenumber of the mode $v_{k}$ and the prime denotes derivative with respect to conformal time $\eta$ defined via $dt = a\,d\eta$.
We see that $z''/z$ introduces a scale in the dynamics: in the sub-Hubble limit $k^{2} \gg z''/z$, the modes oscillate with constant frequency $k$ (they do not feel the curvature of spacetime), while in the super-Hubble limit $k^{2} \ll z''/z$ the modes behave as $v_k \sim z$ (they do feel the curvature), implying that $\mathcal{R}$ is constant for those modes.

Notice that, since the modes $v_{k}$ have a time-dependent frequency, the associated Hamiltonian depends explicitly on time and hence the choice of the vacuum of the theory cannot be done in a time-independent way. Thus, one has to pick an initial time $\eta_{0}$ and define there the vacuum as the lowest energy state. In the limit where $\eta_{0}\to -\infty$ this state is called the Bunch-Davies vacuum~\cite{BunchDavies} and corresponds to the recovery of plane wave solutions in the asymptotic past for sub-Hubble modes (with $k \gg aH$).

From the comoving curvature perturbation one can define the power spectrum as the Fourier transform of its spacetime two-point correlation function~\cite{Mukhanov:1990me}, yielding:
\begin{equation}\label{powerspectrum}
    \mathcal{P}_{\mathcal{R}} =  \frac{k^{3}}{2\pi^{2}} \left|\frac{v_{k}}{z}\right|^{2}.
\end{equation}
This quantifies the contribution to the variance of $\mathcal{R}$ of modes with comoving wavenumber $k$, that is, of quantum zero-point fluctuations. This power spectrum may be evaluated at super-horizon scales, due to constancy of $\mathcal{R}$ for those modes, whence it follows that the power spectrum can be written as:
\begin{equation}
    \mathcal{P}_{\mathcal{R}}(k) = A_{S} \bigl(k / k_{*}\bigr)^{n_{S}-1},
\end{equation}
where $n_{S}$ is the scalar spectral index (or tilt), $A_{S}$ is the scalar amplitude,
and $k_{*}$ is commonly known as the pivot scale from which the power spectrum is measured. The most recent measurements~\cite{Planck:2018vyg} provide the following values at the pivot scale $k_{*} = 0.05 \: \textup{Mpc}^{-1}$:
\begin{equation}\label{PS measurements}
    A_{S} = \left(2.092 \pm 0.034\right) \cdot 10^{-9}, \  n_{S} = 0.9626 \pm 0.0057.
\end{equation}
This means that the power spectrum of fluctuations is nearly scale invariant ($n_{S} \approx 1$), in agreement with the description provided by slow-roll inflation.

\section{Trans-Planckian problem in inflation}\label{TP Inflation}
Despite the great success of the inflationary theory, which we have summarized in Sec.~\ref{Inflation}, this theory also faces several problems. One of them, which is the one that concerns this work, is commonly referred to as the trans-Planckian problem of inflationary cosmology~\cite{BranMart_TransPlanckianProblem}.

In most inflationary models based on an inflaton, the inflationary stage lasts a very large number of $e$-folds. Due to the fact that physical wavenumbers $\kappa = k / a$ at different times are related through $\kappa(t_{1})a(t_{1}) = \kappa(t_{2})a(t_{2})$, one gets that $\kappa(t_{i}) = e^{N} \kappa(t_{f})$. Hence, some of the physical wavenumbers and energies corresponding nowadays to large-scale structures that can be measured in the CMB could be larger (indeed, much larger) than the Planck mass at the beginning of the inflationary stage. 

This is clearly a severe issue, since it implies that the power spectrum of cosmological fluctuations (which is calculated on pure classical gravity) depends as well on high energy Physics. Moreover, the power spectrum we observe today may be altered by any slight modification of Physics above the Planck scale. Therefore, to compute it, it is necessary to be aware of trans-Planckian effects through the evolution of perturbations. However, these effects are yet unknown, and thus the only way to proceed is by introducing reasonable modifications to the theory that try to simulate those effects.

The usual approach to do so is by introducing modified dispersion relations in the equation of motion \eqref{muksaseqn}, as was first done in~\cite{BranMart_TransPlanckianProblem} following the steps of the analog problem in black hole physics~\cite{Jacobson:1999zk}. In this case, due to the spacetime expansion, the analysis is not just an extension of what was done with black holes.
When considering possible trans-Planckian effects, a way of implementing them is by modifying the standard frequency to:
\begin{equation}\label{modifreq}
    \omega^{2}_{F}(\eta) = \bigl[a(\eta)F(\kappa)\bigr]^{2} - z'' / z,
\end{equation}
where $F(\kappa)=F(k/a)$ is some nonlinear function that deviates from the standard linear dispersion relation for physical wavenumbers $\kappa \gg \kappa_c$ and recovers the linear behaviour for $\kappa \ll \kappa_c$, where $\kappa_c$ is some ultraviolet scale (expected to be of the order of the Planck mass). When doing so, a non-Lorentz invariant (in free fall) dispersion relation results, so one must stipulate the reference frame where the dispersion relation is defined.

Since now the dispersion relation is nonlinear, the vacuum cannot be defined as the Bunch-Davies state. Here, we will take the adiabatic approach~\cite{Birrell:1982ix}. The adiabatic vacuum is defined by the positive frequency WKB solution to the Mukhanov-Sasaki equation with modified dispersion relation, appropriately normalized~\cite{BranMart_TransPlanckianProblem}. In this vacuum the modes with $\kappa \ll H$ are in the ground state~\cite{Liddle:1993fq}.

\begin{figure}
    \centering
    \includegraphics[width=0.44\textwidth]{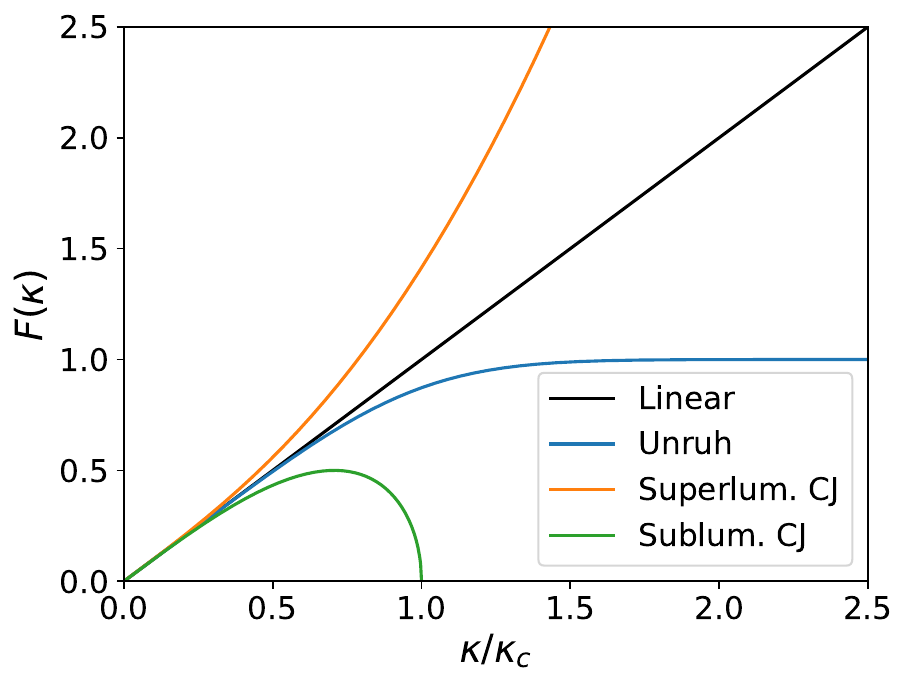}
    \caption{Sketch of the different dispersion relations considered in this work. For the Corley-Jacobson dispersion relation we have considered $m=1$ and $b_m = \pm 1$. }
    \label{Fig. Modified dispersion relations}
\end{figure}

Different modified dispersion relations have been considered until now, some of them shown in Figure~\ref{Fig. Modified dispersion relations}. This includes the so-called Unruh dispersion relation $F_{\textup{U}}$~\cite{Unruh:1994je} or the generalized Corley-Jacobson dispersion relation $F_{\textup{CJ}}$ (introduced in~\cite{BranMart_TransPlanckianProblem} based on the one used in~\cite{Corley:1996ar}):
\begin{align}
          F_{\textup{U}}(\kappa)  &= \kappa_c \tanh{( {\kappa}/{\kappa_c})},\\
          F_{\textup{CJ}}(\kappa) &= \kappa \sqrt{1 + b_{m} ( {\kappa}/{\kappa_c})^{2m}},
\end{align}
where $b_{m}$ reflects the subluminal ($b_{m} < 0$) or superluminal ($b_{m} > 0$) character of $F_{\textup{CJ}}$. In this last family of modified dispersion relations, the nonlinear term must be understood as the first term of a power expansion of a generic dispersion relation; otherwise, one would get (for the subluminal case) pathological behaviour for physical wavenumbers $\kappa > \kappa_c |b_{m}|^{2m}$, where $F_{\textup{CJ}}(\kappa)$ becomes purely  imaginary. Furthermore, in this case the energy may not be bounded from below and the definition of vacuum would not be clear. For these reasons, we will not consider this modified dispersion relation with $b_{m} < 0$ and we will focus only on monotonic dispersion relations. Other modified dispersion relations have been studied in different works ~\cite{Kowalski-Glikman:2000wmg, Lemoine:2001ar, Martin2003}. 

In most works, the robustness of the predictions of inflationary cosmology against trans-Planckian Physics is studied in terms of the adiabaticity coefficient, defined as in~\cite{NiemPar_Adiabaticity} (for an alternative, but equivalent definition see~\cite{BranMart_TransPlanckianProblem}):
\begin{equation}\label{AdiabCoeff}
    \varepsilon(\eta,k) = \bigl|\omega_{F}' / \omega_{F}^{2}\bigr|.
\end{equation}
This coefficient serves as an estimator of how sensitive the power spectrum is to trans-Planckian effects, since a very small value of $\varepsilon$ for trans-Planckian wavenumbers $k$ means that they are very adiabatic in the trans-Planckian regime, and this gives rise to a negligible modification in the power spectrum with respect to the standard prediction, as we will see.

This expression is only valid when the modes are inside the horizon, that is, when $\omega^2 > 0$. In the case $\omega^2 < 0$, the modes fail to be adiabatic as they cross the horizon and instead of oscillating they suffer an exponential amplification. Nevertheless, observable modes are already sub-Planckian when they cross the horizon during slow-roll and thus $\omega^2$ becomes negative when the dispersion relation is the standard one. In particular, we are interested in the value of $\varepsilon(\eta,k)$ during inflation, when a concrete mode is trans-Planckian.  Hence, in its evaluation, the second term in \eqref{modifreq} can be safely neglected, since observable modes are inside the horizon while being trans-Planckian and their dispersion relation dominates over spacetime expansion. Under this approximation, the adiabaticity coefficient can be readily computed:
\begin{equation}\label{AdiabCoeffApprox}
    \varepsilon(\eta,k) \approx \Bigl|\frac{H}{F} - \frac{H\kappa}{F^2} \frac{dF}{d\kappa}\Bigr| = \frac{H}{\kappa_c} \Bigr|\frac{d}{d\kappa}\Bigl(\frac{\kappa \kappa_c}{F}\Bigr)\Bigr|.
\end{equation}
It is easy to see that $\varepsilon$ is bounded by $H/\kappa_c$ for every monotonic dispersion relation. Therefore, one can conclude that as long as the scale separation condition 
\begin{equation}\label{ScaleSeparation}
    H / \kappa_c \ll 1
\end{equation}
is met, the adiabaticity coefficient is $\varepsilon(\eta,k) \ll 1$ for all monotonic modified dispersion relations~\cite{NiemPar_Adiabaticity}, and modifying the standard dispersion relation above the scale $\kappa_c$ will not have an imprint on the power spectrum~\cite{Starobinsky:2001kn}. This is in good agreement with the analytical results obtained in~\cite{BranMart_TransPlanckianProblem,BranMart_Adiabaticity}. Physically, this can be seen as if the modes with $\kappa > \kappa_c$ (which are affected by modified dispersion relations) have enough time to adapt   to the standard vacuum solution provided that their evolution is adiabatic, since $F\geq H$ for those $\kappa$ values that satisfy the adiabaticity condition. Moreover, scale separation \eqref{ScaleSeparation} is satisfied during slow-roll, as long as $\kappa_c \sim m_{\textup{Pl}}$, since then $H/\kappa_c \lesssim 10^{-5}$.

\section{Loop Quantum Cosmology and the trans-Planckian problem}\label{LQC}

\subsection{Loop Quantum Cosmology: Background}\label{LQC Background}

Let us first summarize the evolution of the homogeneous background obtained in LQC (for a complete review of this theory and its derivation, see~\cite{AshtekarLQC_Review,Agullo:2016tjh}).

The quantum geometry effects that this theory introduces allow to remove the classical big bang singularity of FLRW models, replacing it by a quantum bounce, hence giving rise to a well-defined background evolution of the Universe. In fact, LQC leads to a family of semi-classical states that follow well-defined trajectories. These trajectories correspond to an effective dynamics encoded in the following modified Einstein equation with quantum corrections~\cite{AshtekarLQC_Review}:
    \begin{align}
        H^{2}_{\textup{LQC}} = \frac{8\pi }{3 m_{\textup{Pl}}^2} \rho \bigl(1 - {\rho}/{\rho_{*}}\bigr) \label{friedeqnLQC},
    \end{align}
where $\rho_{*}$ is a critical density of the order of the Planck density, which is usually taken to be $\rho_{*}~ \approx~ 0.41 m_{\textup{Pl}}^4$ due to geometrical arguments~\cite{Ashtekar:2009vc}.
We see that, indeed, equation \eqref{friedeqnLQC} leads to a bounce, where the Hubble parameter vanishes, when $\rho = \rho_{*}$. This is a feature of LQC that is not present in General Relativity (GR) and guarantees that physical quantities (such as the energy density or the Ricci scalar) that diverge in GR are bounded in LQC. Moreover, the term $\rho/\rho_{*}$ is negligible a few Planck seconds after the bounce, so one recovers GR soon after it. 

\begin{figure}
    \centering
    \includegraphics[width=0.45\textwidth]{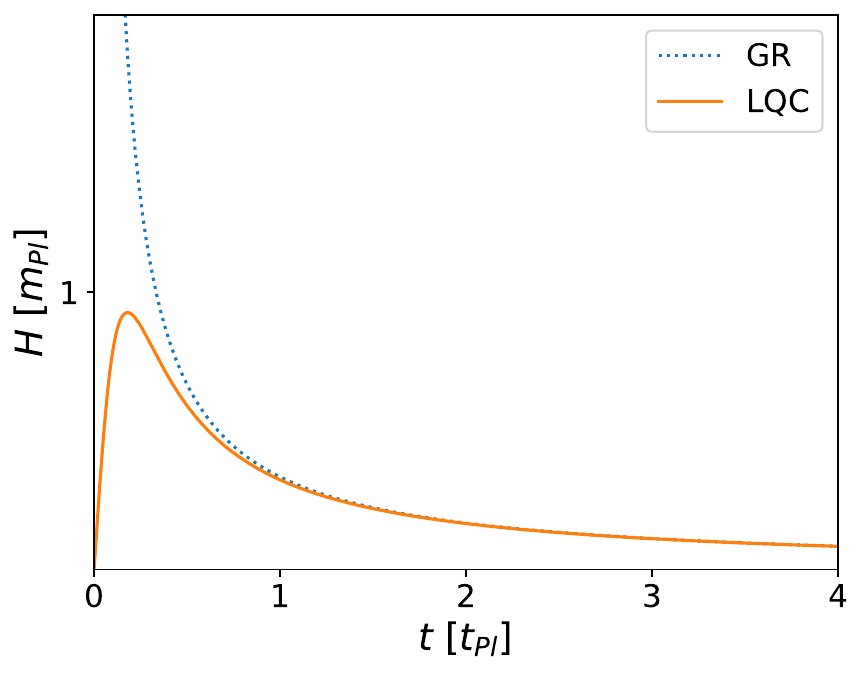}
    \caption{Hubble parameter for a free scalar field in GR and LQC (with $\rho_{*} = 0.41 m_{\textup{Pl}}^4$). $H_{\textup{LQC}}$ is always bounded and the GR behaviour is restored soon after the bounce. For larger (smaller) values of $\rho_{*}$, GR is recovered sooner (later) and the maximum of $H_{\textup{LQC}}$ is different.}
    \label{Fig. Hubble parameter GR vs LQC}
\end{figure}

In LQC, the scalar field can also be described as a perfect fluid obeying \eqref{backgroundeqns}, but now with the Hubble parameter given by \eqref{friedeqnLQC}. This set of equations is analytically intractable except for the free scalar field case (and a few others). In this case where $V(\phi) = 0$ the analytical solution in terms of cosmological time $t$ is (from now on, the subscript LQC will be omitted unless necessary for comparison with GR):
\begin{align}
    a(t)&=\left[\left(\frac{t}{t_{*}}\right)^2 + 1\right]^{1/6}, \qquad\qquad\qquad H(t)= \frac{t}{3t_{*}^2 a^6(t)} ,  \nonumber \\
    \phi(t)&=\phi(0) + \frac{m_{\textup{Pl}}}{\sqrt{12\pi }} \arcsinh{\left(\frac{t}{t_{*}}\right)}, \qquad \rho(t) = \frac{\rho_{*}}{a^{6}(t)},
   \end{align}
where $t_{*} = m_{\textup{Pl}}/\left(24\pi   \rho_{*}\right)^{1/2}$ is the time when the Hubble parameter achieves its maximum value $H(t_{*}) = \sqrt{2\pi \rho_{*} / (3m_{\textup{Pl}}^2)}$ and we have chosen as normalization the scale factor at the  bounce $a(0) = 1$. For $\rho_{*} = 0.41 m_{\textup{Pl}}^4$ this maximum is $H(t_{*}) \approx 0.93\,m_{\textup{Pl}}$. The Hubble parameter is depicted in Figure~\ref{Fig. Hubble parameter GR vs LQC}, in comparison with the classical GR behaviour.
Similar plots can be portrayed for the other background variables, leading to the conclusion that LQC cures the initial singularity standard cosmology displays for a free scalar field and enables to define a pre-inflationary dynamics. 

In addition to solving the singularity problem, LQC also provides natural initial conditions that ensure the inflationary paradigm with the required number of $e$-folds for suitable potentials~\cite{AshtekarSlowRoll,SinghVandersloot}, after a very short phase of super-inflation takes place. However, to make the inflationary stage happen we need a non-vanishing potential, which requires numerical treatment. In this paper, we will use both the quadratic and Starobinsky family of potentials:
    \begin{align}
        V_2(\phi) &=  \frac{1}{2} m^{2}\phi^{2} , \\
        V_\textup{S}(\phi) &= V_{0} \left( 1 - \exp{\left\lbrace -\sqrt{\frac{16\pi }{3 m_{\textup{Pl}}^2}} \phi\right\rbrace} \right)^2
        \label{StarPotential},
    \end{align}
where $m$ and $V_{0}$ are the parameters that determine the concrete potential of these families.
We will mainly focus on the Starobinsky potential model, which is favoured by data in the case of standard cosmology~\cite{Planck:2018jri}, and based on the conclusions obtained we will briefly review a quadratic potential model.

In Sec.~\ref{Numerical Results} we will see that, as long as there is kinetic dominance at the bounce (which is the case we will be interested in), the background evolution is that of the free field case from the bounce until a short time after the maximum of $H$ and, after some time, the Hubble parameter goes to a non-vanishing constant, thus causing inflation (which implies a dominance of the potential). Moreover, numerical simulations show that under this assumption the background evolution for non-vanishing potentials at early times does not depend strongly on the choice of initial conditions or on the concrete shape of the potential, and can be well-described by the zero potential solution~\cite{Zhu:2017jew}; thus, the maximum of $H$ and the time when it is reached depend only on the value of~$\rho_{*}$.

\subsection{Loop Quantum Cosmology: Perturbations}\label{LQC Perturbations}

We next consider cosmological perturbation theory in an LQC effective cosmological background. This would enhance our knowledge of the Universe evolution and enable us to connect Planck era with observable quantities. In this case, in addition to possible effects of trans-Planckian Physics in ultraviolet modes (as those described in Sec.~\ref{TP Inflation}), modifications on infrared modes could arise as well due to a different evolution of the background: since these modes feel the curvature at the bounce (their wavelength is bigger than the curvature radius of the Universe at that time) they can evolve to an excited state. Indeed, observable modes could in principle exit and re-enter the curvature radius before inflation takes place, thus reaching the onset of inflation in a vacuum other than Bunch-Davies.

Current techniques only allow us to obtain these infrared corrections to the propagation of the perturbations, as consequence of the LQC quantisation of the background. Different strategies have been followed to accomplish this task (see~\cite{Agullo_SummaryPertLQC} for a review), namely, the so-called dressed metric approach~\cite{AshtekarLewandowski_DressedMetric, AgulloAshtekar1_DressedMetric, AgulloAshtekar2_DressedMetric, AgulloAshtekar3_DressedMetric}, the deformed algebra approach~\cite{Bojowald:2008jv}, the hybrid quantisation~\cite{MenaMerceNavascues, MenaOlmedoMikel}, or the separate Universe loop quantisation~\cite{Wilson-Ewing:2015sfx}. For our purposes we will consider hybrid LQC. This is because the separate Universe approach is only valid for infrared modes, whereas the deformed algebra approach seems to be incompatible with observations~\cite{Bolliet:2015raa}. On the other hand, hybrid and dressed metric approaches are based mainly on the same grounds, but their equations of motion for perturbations are not the same, due to the different ways of quantising.

In particular, in hybrid LQC, when neglecting the back-reaction of the perturbation modes, the equation for scalar perturbation modes has the same structure as \eqref{muksaseqn} but with a time-dependent frequency~\cite{Mena_TimeDependentMasses}
\begin{equation}\label{muksaseqnLQC}
   \omega_{k}^{2}(\eta) = k^{2} + s(\eta),\qquad
    s = - \frac{4\pi }{3 m_{\textup{Pl}}^2} a^{2} \left(\rho - 3p\right) + a^2u,
\end{equation}
where $u$ is an effective potential given by 
\begin{equation}
    u =   V''(\phi)  +  \frac{48\pi}{m_{\textup{Pl}}^2} V(\phi)\bigg(1-\frac{V(\phi)}{\rho}\bigg) +  6 \frac{a'\phi'}{a^{3}\rho}V'(\phi).
\end{equation}
In the classical limit where the effective dynamics reduces to GR, that is, when $\rho_{*} \to \infty$, we recover the classical Mukhanov-Sasaki equation \eqref{muksaseqn}. We also note that during kinetic dominance the above function $s(t)$ is positive, unlike the analog function in the dressed metric approach~\cite{Mena_TimeDependentMasses}. This is the reason why we will choose to work within hybrid LQC, avoiding further problems related to the presence of modes with complex frequencies at the bounce, as we will see later in Sec.~\ref{Numerical Results Perturbations}.

\subsection{Trans-Planckian problem in Loop Quantum Cosmology}

Now that we have the explicit form of the infrared corrections to the perturbations coming from LQC, we may wonder how and why LQC and Planck-scale Physics may affect the power spectrum in the ultraviolet, in analogy with Sec.~\ref{TP Inflation} that focused solely on inflation. An extra drawback appears in the context of LQC, as was mentioned before, namely, the relevance of the background in modes which are sensitive to the curvature. Moreover, LQC introduces a scale as it already happened in inflation. In this case, this scale results from the competition between the physical wavenumber of the modes and the Ricci scalar in the Mukhanov-Sasaki equation (equivalently, between the physical wavelength and the curvature radius). The discrimination between which modes feel the curvature and which do not is precisely the value of the Ricci scalar at the bounce, where it is maximum, and constitutes the characteristic energy scale of LQC:
\begin{equation}\label{LQCscale}
    \kappa_{\textup{LQC}} = \sqrt{R(0)/6} = (\sqrt{3}t_{*})^{-1} \approx 3.21\,m_{\textup{Pl}}.
\end{equation}
Modes with $\kappa \lesssim \kappa_{\textup{LQC}}$ will be the ones which feel the curvature since they have a wavelength longer than the LQC wavelength. Consequently, they exit and enter the curvature radius in the bouncing stage, before they exit again in inflation, which can have an imprint on the power spectrum due to background effects.

Another central issue is the choice of initial conditions for the perturbations. There are different alternatives, namely, setting them far away before the bounce (in the contracting branch) or at the bounce itself. The former allows to set initial conditions at some point where effects of modified dispersion relations disappear, but requires fixing them at the concrete points where $s(\eta) = 0$ for stability~\cite{FrancesesMonomialPotentials}. On the other hand, the latter, although facing the problem that the bounce is the most affected region by LQC, at least guarantees that all perturbations behave there as harmonic oscillators with real frequencies. However, it is only the infrared part of the spectrum that is sensitive to the choice of initial conditions. For a review of this topic and the computation of the power spectrum for several monomial potentials with the standard dispersion relation, see~\cite{FrancesesMonomialPotentials}.  

The methodology that we will follow is the same as before. One includes in \eqref{muksaseqnLQC} modified dispersion relations (replacing $k^{2}$ by $a^{2}F^{2}$) and analyses how fair the assumption of WKB solution is by means of the pertinent adiabaticity coefficient when modes are trans-Planckian.

To our knowledge, not much work has been done in the trans-Planckian problem in LQC. In particular, only in~\cite{Francesesmdr} modified dispersion relations were considered within the framework of LQC. In that
work, the emphasis was placed in computing the power spectrum in different approaches and comparing with the standard scenario, rather than understanding the possible modification and its origin qualitatively. The main result was that the power spectrum is modified when considering modified dispersion relations and that this change depends on the concrete value of the ultraviolet scale $\kappa_c$, possibly leading to a power spectrum with oscillations or enhancement in its ultraviolet sector, thus making relevant the trans-Planckian effects. This opens the question of whether such effects affect or not the observable window.

The calculation of the adiabaticity coefficient can be carried out easily for the kinetic dominated regime close to the bounce, as it is well approximated by the free scalar field case, where the explicit background solution is known and $u=0$, yielding $s_{0}(\eta) = 8\pi  \rho a^{2} / (3m_{\textup{Pl}}^2)$, with $\rho=\dot\phi^2/2$. In this case, we have:
\begin{equation}\label{AdiabCoeffZeroLQC}
    \varepsilon_{0}(\eta,k) = \frac{H F^3}{\kappa_c \mathcal{W}_{0}^{3}} \Bigl| \frac{d}{d\kappa} \Bigl(\frac{\kappa \kappa_c}{F}\Bigr) - \frac{16\pi }{3m_{\textup{Pl}}^2} \frac{\rho \kappa_c}{F^{3}} \Bigr|,
\end{equation}
where
\begin{equation}
    \mathcal{W}_{0} = \frac{\omega}{a} = \sqrt{F^{2} + \frac{8\pi }{3m_{\textup{Pl}}^2} \rho}.
\end{equation}
Let us qualitatively analyse this adiabaticity coefficient. 

First, notice that it is proportional to $H$, so that the adiabaticity condition $\varepsilon_0 \ll 1$ for trans-Planckian modes again follows from the condition $H / \kappa_c \ll 1$. According to \eqref{friedeqnLQC}, we have that $\varepsilon_0(\eta,k) \propto (1-\rho / \rho_*)^{1/2}$. Hence, as long as $F$ is not too steep, all the modes satisfy the adiabaticity condition very close to the bounce (where $\rho \approx \rho_{*}$). Moreover, $\varepsilon_0(\eta,k)$ is exactly zero at the bounce. Therefore, all the modes can be set up~at the bounce in their adiabatic vacuum, regardless of the specific modified dispersion relation and the physical wavenumber of the modes.

Second, when the evolution of perturbations is fully determined by its dispersion relation (that is to say, when $\mathcal{W}_0 \approx F$), we recover the inflationary result \eqref{AdiabCoeffApprox}.

A similar result to \eqref{AdiabCoeffZeroLQC} can be derived when the potential cannot be ignored, with an extra term that depends only on the specific potential and on the background, but not on the modified dispersion relation. In this case, we do not have an analytical solution for the background evolution, so we will need to compute $\varepsilon(\eta,k)$ numerically introducing \eqref{muksaseqnLQC} in the definition \eqref{AdiabCoeff}.

\section{Numerical results}\label{Numerical Results}

\subsection{Background dynamics}

Let us start with the evolution of the background solution, which is necessary to analyse the evolution of primordial perturbations. As stated in the previous section, we will concentrate on the Starobinsky potential model. We have taken $V_0 = 1.77 \cdot 10^{-13}\,m_{\textup{Pl}}^4$ in order to consider a model in LQC that is compatible with observations for the standard dispersion relation~\cite{Bonga:2015kaa}.

The background evolution is completely characterized by the initial condition $\phi(0)$, since $\dot{\phi}(0)$ is fixed by $\rho(0) = \rho_*$, as $H = 0$ at the bounce. We fix the normalization of the scale factor at the bounce to be $a(0) = 1$, as in the zero-potential example above.

We have considered two different background evolutions which are compatible with observations in the CMB for the standard dispersion relation.
The first one is based on~\cite{AshtekarValues}. This work, employing the dressed metric approach, fixes initial conditions according to first principles that aim to link quantum geometry and Heisenberg uncertainties in the Planck epoch with late time Physics. This leads to a background evolution given by $\phi(0) = -1.42\,m_{\textup{Pl}}$, which produces $141.30$ $e$-folds from the bounce until today, as found in \cite{AshtekarValues}. We will adopt this background initial condition in the hybrid LQC approach, and refer to it as model A.\footnote{As mentioned, we work with the hybrid approach because of the drawbacks of the dressed metric approach. Nevertheless, the initial conditions we are concerned with at the moment refer only to the background dynamics, which is the same in both of them. When analysing the results, we comment on what would happen if we had followed the dressed metric procedure, which further corroborates our choice.} The second one is based on~\cite{MRJ}. This work, using the hybrid LQC approach, considers a family of vacua for cosmological perturbations that lead to a power spectrum with infrared exponential suppression. The scale where this suppression occurs is fixed so that it is preferred by Planck data, via a Bayesian analysis. In this case, $\phi(0) = -1.46\,m_{\textup{Pl}}$, which produces $131.83$ $e$-folds from the bounce until today.\footnote{\label{fn:calculations}In \cite{MRJ} only the number of $e$-folds of inflation is cited. We have computed the total number of $e$-folds by comparing the physical wavenumber of a reference scale today and at the bounce: $N =~\ln(\kappa(\text{bounce})/\kappa(\text{today}))$. Specifically, we take the pivot scale of the PLANCK collaboration $\kappa_p(\text{today}) = 0.05\,\textup{Mpc}^{-1}$ which is defined as the scale at which the primordial power spectrum has the amplitude $A_s = e^{3.047}\cdot 10^{-10}$ \cite{Planck:2018vyg}. By identifying the value of $\kappa_p$ at the end of inflation from the primordial power spectrum, we track it back to the bounce and determine $\kappa_p(\text{bounce}) \simeq 0.05\,\ell_\textup{{pl}}^{-1}$.} We will refer to this as model B.

This difference in the total number of $e$-folds is related to the different expansions that the Universe suffers during its evolution. As we will see later, whereas the number of $e$-folds during the pre-inflationary period is almost the same for both models,  model A has more $e$-folds of inflation than model B.

\begin{figure}\centering
        \includegraphics[width=0.44\textwidth]{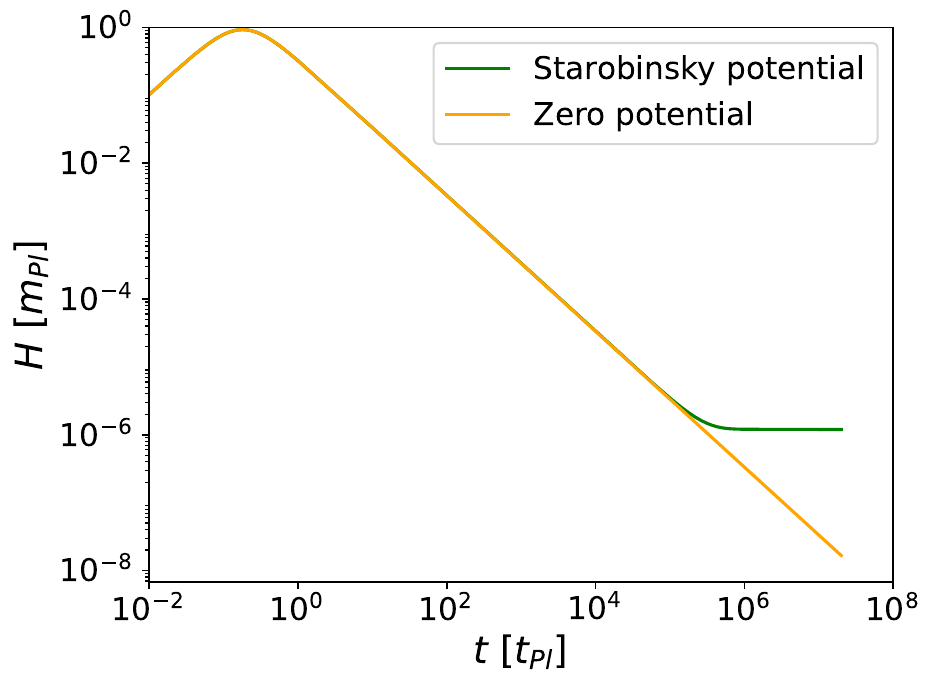}
        \includegraphics[width=0.44\textwidth]{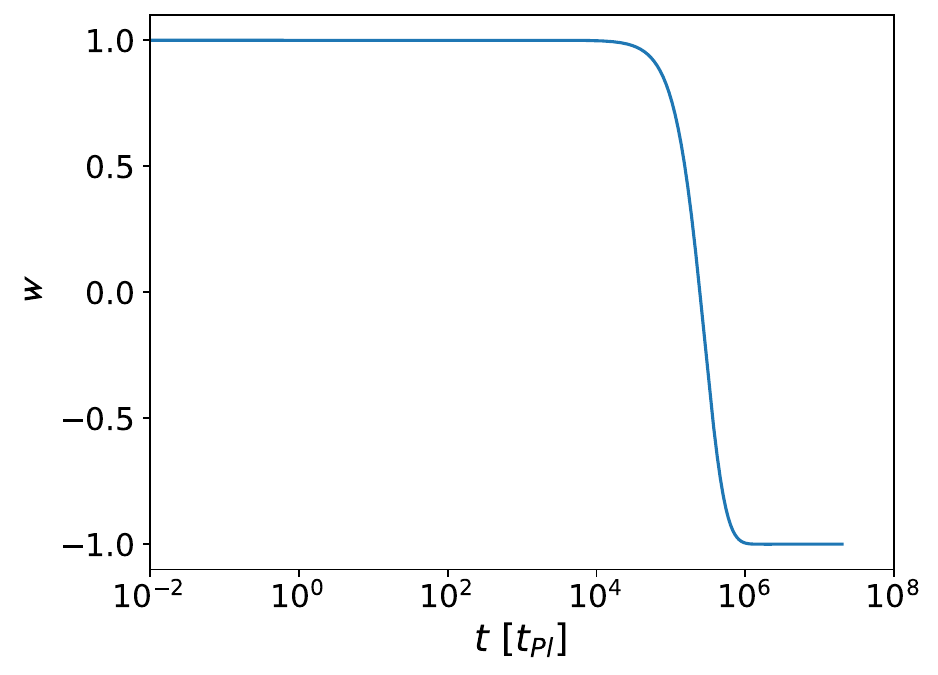}
  \caption{Background evolution for model A. 
  Upper  panel: Hubble parameter under the Starobinsky potential, compared to the zero-potential case.  Lower  panel: Parameter of state of the inflaton.  Inflation begins when $w$ becomes smaller than~$-1/3$. Comparing with the upper panel, one sees that the potential only becomes relevant close to the onset of inflation. Model B results in similar curves for the range plotted in these figures. The differences arise later during the inflationary period, as evidenced by Table~\ref{Table. efolds Star}.
}
\label{Fig. Background dynamics}
\end{figure}

The simulations have been done in cosmological time~$t$, and run up to $t\sim 10^{7}\,t_{\textup{Pl}}$ in order to reach inflation. The background evolution is depicted in Figure~\ref{Fig. Background dynamics} for model~A. As it can be seen on the upper panel, the Hubble parameter evolves near the bounce as if the scalar field  were  free and, around $t \sim 10^{5}\,t_{\textup{Pl}}$, the Starobinsky and zero-potential lines begin to differentiate. From this moment on, $H(t)$ is roughly constant, which means that the scale factor there grows exponentially and slow-roll inflation  takes  place. To get further insight, we have also plotted on the lower panel of Figure~\ref{Fig. Background dynamics} the parameter of state, defined as
\begin{equation}
    w(\phi) = \frac{p}{\rho} = \frac{\dot{\phi}^{2}/2 - V(\phi)}{\dot{\phi}^{2}/2 + V(\phi)}.
\end{equation}
Near the bounce there is kinetic dominance and hence $w \approx 1$, whilst during slow-roll inflation $w \approx -1$, thus confirming potential dominance. After inflation, since the scalar field begins to oscillate, so will   $w$.

Similar portraits result when depicting the background solution of model B.  The difference arises in the number of $e$-folds of the inflationary period\footnote{Here, the start and end of inflation is determined by the condition $w = - 1/3$, that is to say, when the Universe begins or stops its accelerated expansion, respectively.}. Results for both models are displayed in Table~\ref{Table. efolds Star}. As we can see, the pre-inflationary stage lasts almost the same   number of $e$-folds  independently of the model we are using, that is to say, independently of the initial conditions considered. This is because the initial conditions considered reproduce a kinetic dominance bouncing stage, in which case the choice of initial conditions does not strongly affect the background evolution at early times, as we stated at the end of Sec.~\ref{LQC Background}. On the other hand, the number of $e$-folds during inflation is different depending on the model. In particular, inflation lasts almost $11$ $e$-folds more in model A than in model B. In summary, the initial condition at the bounce $\phi(0)$ only affects the number of $e$-folds of inflation.

\begin{table}
    \centering
    \begin{tabular}{|c|c|c|c|c|}
        \cline{2-5}
        \multicolumn{1}{c|}{}
        & \multicolumn{2}{c|}{Starobinsky} & \multicolumn{2}{c|}{Quadratic} \\
        \cline{2-5}
        \multicolumn{1}{c|}{\multirow{2}{*}{}} & Model A & Model B & Model A & Model B \\
        \hline
        $N_{\textup{preinf}}$ & 4.862 & 4.863 & 4.228 & 4.237 \\
        $N_{\textup{inf}}$ & 72.245 & 61.325 & 68.109 & 64.512 \\
        \hline
    \end{tabular}
    \caption{Number of $e$-folds for both models with the Starobinsky and quadratic potentials (commented on at the end of Sec.~\ref{Numerical Results Perturbations}). Model A has more inflation than model B, but the pre-inflationary stage lasts essentially the same   number of $e$-folds.}
    \label{Table. efolds Star}
\end{table}

\subsection{Perturbations}
\label{Numerical Results Perturbations}

Let us now focus on the dynamics of the perturbations in each model, so that we can determine whether there is a trans-Planckian problem in them. As described in the previous section, such  evaluation will rely on the adiabaticity coefficient of the different modes while  they are trans-Planckian. Concretely, the problem arises when observable modes (today) were trans-Planckian at some point in the past while not being adiabatic, as this would indicate that the primordial power spectra are very sensitive to trans-Planckian Physics. Thus, we first need to determine whether the observable modes were ever trans-Planckian in these models. This requires choosing a specific value for the ultraviolet scale~$\kappa_c$. Although other choices could be made, we have chosen $\kappa_c = \kappa_{\textup{LQC}}$ for both models, since it seems natural that the ultraviolet modifications arise when the energy scale of LQC is dominant. We also need to track the observable window measured in the CMB. 
In the case of model A we have $141.30$ $e$-folds from the bounce until today~\cite{AshtekarValues}, which means that the observable window today, \mbox{$\kappa_{\textup{today}} \in [10^{-4}\,\textup{Mpc}^{-1}, 0.5\,\textup{Mpc}^{-1}]$}  \cite{Planck:2018jri},   at the bounce   becomes \mbox{$\kappa(0) \in [1.2\,m_{\textup{Pl}}, 6.1 \cdot 10^{3}\,m_{\textup{Pl}}]$}. Meanwhile, for model B, since there are $131.83$ $e$-folds from the bounce until today, the observable window at the bounce is \mbox{$\kappa(0) \in [9.3 \cdot 10^{-5}\,m_{\textup{Pl}}, 4.7 \cdot 10^{-1}\,m_{\textup{Pl}}]$}.

\begin{figure}
        \centering        \includegraphics[width=0.44\textwidth]{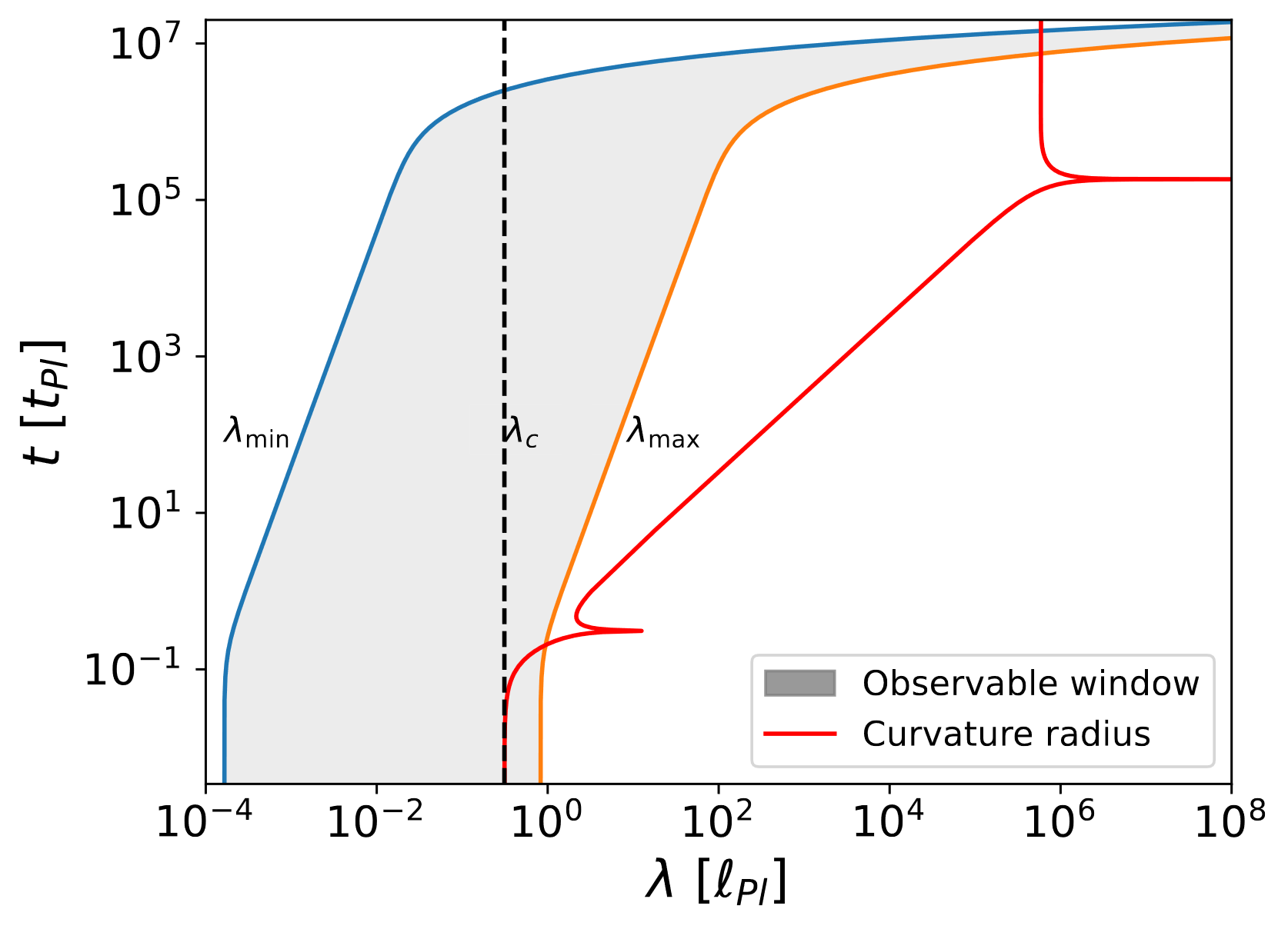}
   \includegraphics[width=0.44\textwidth]{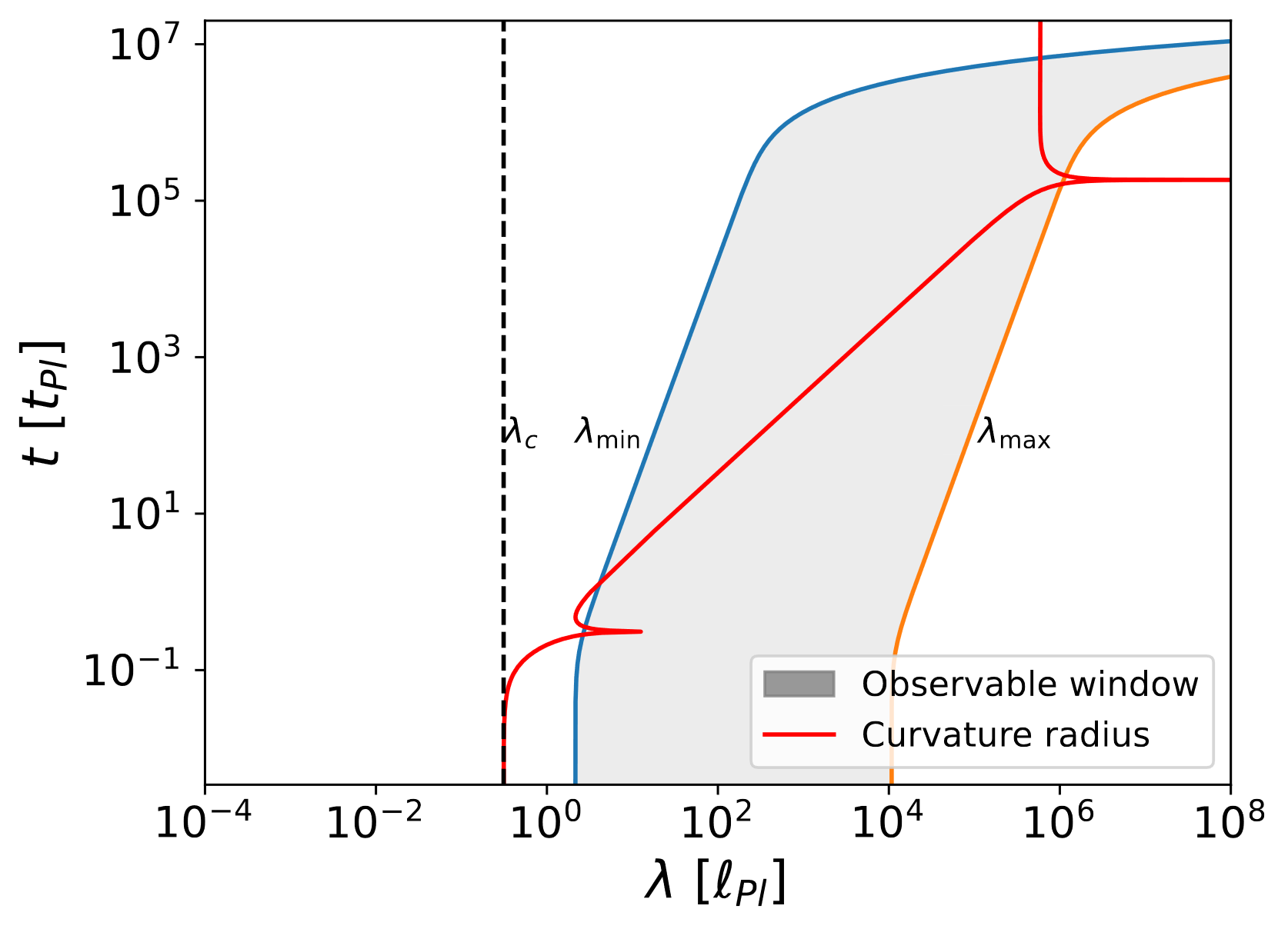}
\caption{Evolution of the observable window (shaded region) from the bounce to inflation.  Upper panel: Model A. A big part of the window is trans-Planckian during the evolution (even at the onset of inflation) before exiting the horizon. Lower panel: Model B. The whole window is always above the ultraviolet scale $\lambda_{c}$.}
\label{Fig. Observable Window}
\end{figure}

We have evolved these windows from the bounce to the end of the simulation along with the curvature radius. The results are shown in Figure~\ref{Fig. Observable Window} in terms of physical wavelengths, where the ultraviolet scale $\lambda_c = 1/\kappa_c \approx 0.31 \ell_{\textup{Pl}}$ is also depicted.

Focusing first on model A (upper panel), we see that the observable window presents the two kinds of modes that we have been discussing. On the one hand, it contains modes whose physical wavelength is below $\lambda_c$, not only at the bounce, but also at the onset of inflation, which means they are trans-Planckian. On the other hand, some modes are outside the curvature radius at the bouncing epoch and thus  are affected by the dynamics of the background (carrying the effects of LQC). These most infrared modes of the observable window will then have an imprint on the power spectrum due to LQC effects, reflected in the loss of near-scale-invariance in that sector~\cite{Agullo_SummaryPertLQC}. Notice as well that all the modes cross the horizon during inflation  when they are no longer trans-Planckian, that is to say, when their physical wavelength is well above the ultraviolet scale $\lambda_c$.

On the other hand, for model B the whole observable window in the lower panel of Figure~\ref{Fig. Observable Window} always has a wavelength greater than the ultraviolet scale $\lambda_{c}$, which means that all the modes are sensitive to the curvature in the pre-inflationary epoch but are never trans-Planckian. This drastic difference with respect to model A has to do with the different number of $e$-folds that both models predict in order to be compatible with observations for the standard dispersion relation. We note that, even though all of these modes are indeed sensitive to the background dynamics in the pre-inflationary epoch in this model, the vacuum chosen in~\cite{MRJ} is such that departure from near-scale-invariance in the primordial power spectrum only occurs for the most infrared observable modes. This leads to predictions that are compatible and indeed preferred by the data.

\subsection{Adiabaticity coefficient}

It is now time to compute the adiabaticity coefficient for the different modified dispersion relations that affect the ultraviolet sector while the modes are trans-Planckian. This will allow us to determine if the ultraviolet part of the observed power spectrum is robust against trans-Planckian effects in these models within LQC.

We have to recall here that model A is taken from a context where the dressed metric approach is used to analyse primordial perturbations. In this paper, we only use the initial conditions and the background solution of this model, and we apply it to hybrid LQC. Otherwise, if the dressed metric approach were to be used, according to its expression for the time-dependent frequency (see~\cite{Mena_TimeDependentMasses}), the time-dependent mass $s$ (analog to  \eqref{muksaseqnLQC}) would be negative at the bounce and equal to $s(t=0) = -8\pi \rho_{*}/m_{\textup{Pl}}^2 = - \kappa_{c}^{2}$, where we have ignored the potential contribution completely. Therefore, at the bounce we would have $\omega_{k}^{2}(t = 0) = k^2 - \kappa_{c}^{2}$ for the standard dispersion relation, so that modes with $k < \kappa_{c}$ at the bounce (more infrared than $\kappa_{c}$) would have a complex time-dependent frequency ($\omega^2 < 0$). This reveals that in the dressed metric approach the adiabaticity for these modes is completely lost in the trans-Planckian regime, and that such a framework suffers from a severe trans-Planckian problem.

This is no longer the case in hybrid LQC. According to \eqref{muksaseqnLQC} at the bounce we  have \mbox{$s(t=0) = 8\pi   \rho_{*} / (3m_{\textup{Pl}}^2) = \kappa_{c}^{2} / 3$}, making the time-dependent frequency always positive for every monotonic dispersion relation. Indeed, $s$ is always positive in hybrid LQC~\cite{Mena_TimeDependentMasses} until we reach inflation, making it possible to perform an adiabatic analysis for all the modes while trans-Planckian effects  may be important.

\begin{figure}
        \centering
        \includegraphics[width=.44\textwidth]{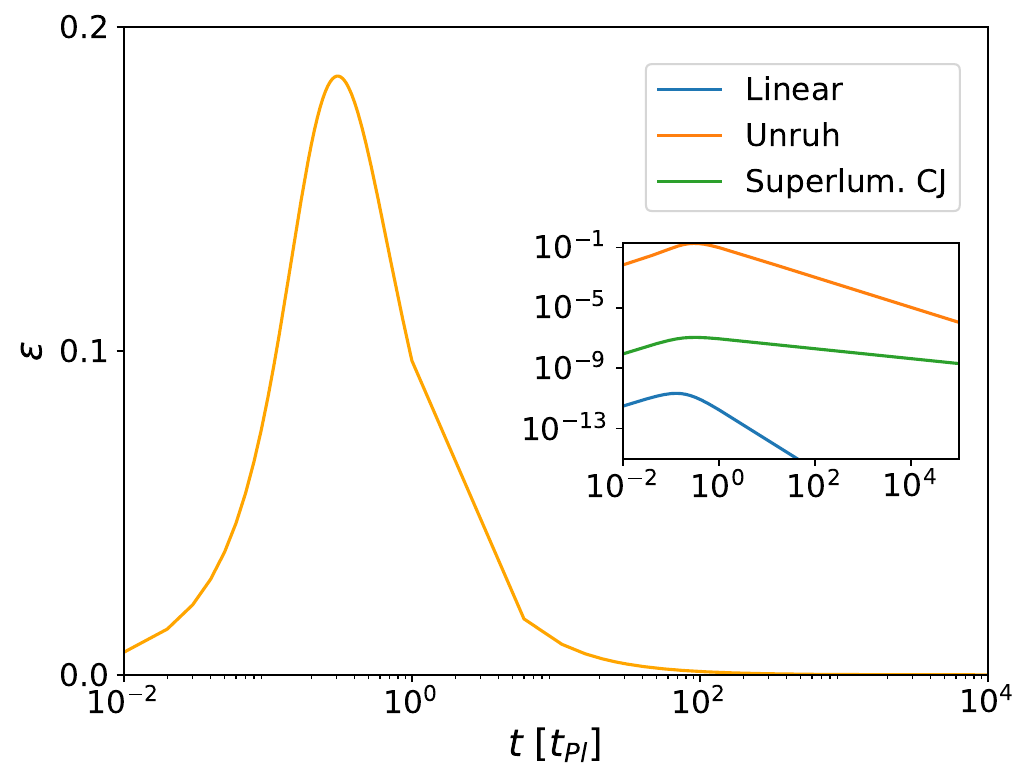}
        \includegraphics[width=.44\textwidth]{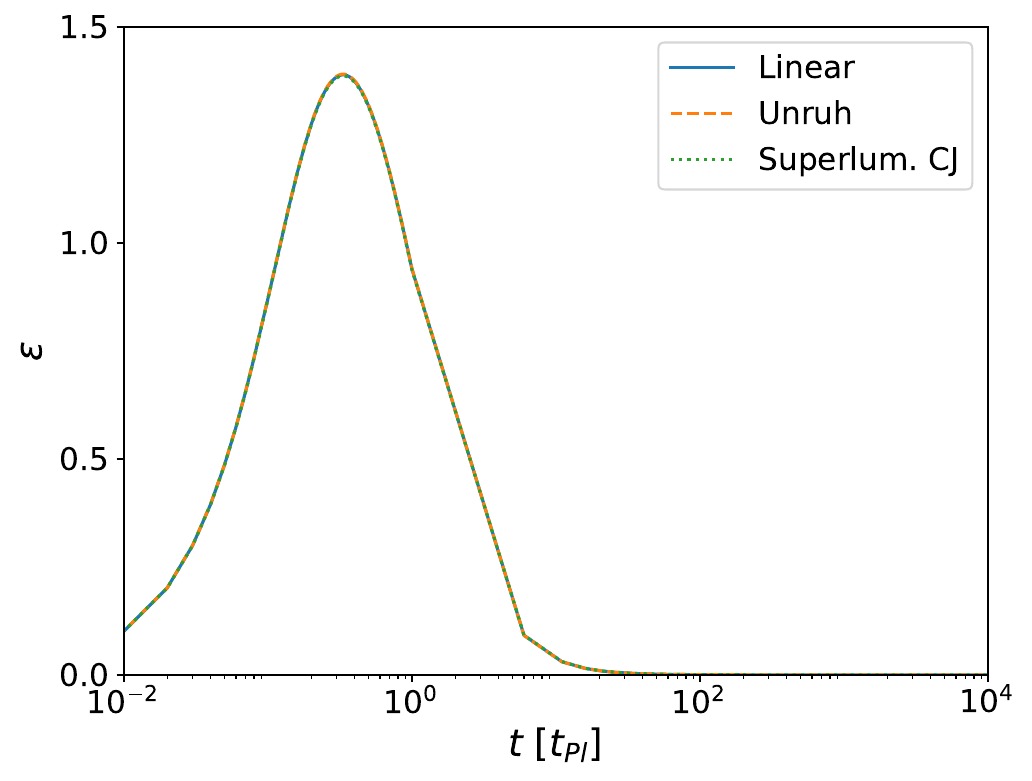}
\caption{Adiabaticity coefficient for the different dispersion relations considered in this work. All the lines depicted tend to zero at the bounce ($t=0$). Upper panel: Model~A. The adiabaticity coefficient is different depending on the modified dispersion relation. In particular, it becomes non-negligible after the bounce for the Unruh dispersion relation, whereas it remains essentially zero for the linear and superluminal Corley-Jacobson  dispersion, as evidenced by the inset. Lower panel: Model~B. The adiabaticity coefficient is the same regardless of the modified dispersion relation  since the three curves overlap.}
\label{Fig. Adiabaticiy coefficient LQC}
\end{figure}

We have calculated the adiabaticity coefficient $\varepsilon(\eta,k)$ both for the Unruh and superluminal Corley-Jacobson modified dispersion relations, as well as for the linear one. The results are depicted in Figure~\ref{Fig. Adiabaticiy coefficient LQC} for both models. We have only represented the ultraviolet endpoint of the observable window, since it is the most sensitive mode to modified dispersion relations among the observable modes.

We now summarize the main results that follow from the analysis of these graphics.  Firstly, the adiabaticity coefficient is noticeably different depending on the model we are using: for model A, $\varepsilon$ is always below one despite the modified dispersion relation we are considering, whereas model B presents an adiabaticity coefficient which is above one at certain points of the cosmological evolution. Another thing to notice is that, regardless of the model, the adiabaticity coefficient achieves its maximum around the same time ($t\sim 10^{-1} t_{\textup{Pl}}$) which is around the   time  at which  the Hubble parameter achieves its maximum. This means that, in the pre-inflationary epoch, modes are the least adiabatic around the peak of~$H$. This reinforces the idea that scale separation \eqref{ScaleSeparation} is required in order to have $\varepsilon \ll 1$, which we discussed in Section~\ref{TP Inflation}.  Indeed, we have $H / \kappa_c \approx 0.29$ at the maximum of $H$, which is not so far below $1$, so it is not surprising that in this region the adiabaticity coefficient  is not negligible.

Let us focus on the upper panel of Figure~\ref{Fig. Adiabaticiy coefficient LQC} corresponding to the results for model A. We see that the adiabaticity coefficient of the shortest wavelength mode differs depending on the modified dispersion relation. In particular, for the linear and superluminal Corley-Jacobson dispersion relations $\varepsilon$ remains far below one at every instant while the mode is trans-Planckian. On the other hand, with the Unruh dispersion relation $\varepsilon$ is non-negligible at the bounce and afterwards, reaching a value of almost $0.2$ for the shortest wavelength mode. This implies that the whole observable window is non-adiabatic right after the bounce,  when most of the modes are trans-Planckian (according to Figure~\ref{Fig. Observable Window}, upper panel). We therefore conclude that model A presents a trans-Planckian problem: as a result of our analysis we can foresee that the power spectrum depends on the particular modified dispersion relation, suffering non-negligible departures from near-scale-invariance in the most ultraviolet observable modes. However, we note that this dependence is soft (as $\varepsilon<1$), and that for some dispersion relations all the observable modes remain adiabatic throughout the whole pre-inflationary epoch. One may argue that this is a soft trans-Planckian problem, especially when compared to the same initial conditions within the dressed metric approach, since in the latter case $\varepsilon$ diverges for some observable modes close to the bounce. 

Now let us look at the lower panel of Figure~\ref{Fig. Adiabaticiy coefficient LQC}, representing the results obtained for model B. Noticeably, the adiabaticity coefficient is larger than one for an extended period of time. In contrast with model A, there is a period of the pre-inflationary evolution when (at least some of) the observable modes are categorically non-adiabatic. However, this is not a problem in this model since the modes are not trans-Planckian in this epoch. In fact, it is just as in standard cosmology when the modes are not adiabatic as they cross the horizon during inflation. We see that for the three dispersion relations considered in this work the adiabaticity coefficient is the same and the three different curves overlap. This is because the whole observable window is sub-Planckian in this epoch (according to Figure~\ref{Fig. Observable Window}, lower panel), and therefore corresponds to the approximately linear regime of all dispersion relations. Thus, the observable window of the primordial power spectra will not be affected by the choice of dispersion relation.

\subsection{Quadratic potential}

To conclude this section, let us dedicate a comment to the quadratic potential case. For this potential, we have to take $m = 1.21 \cdot 10^{-6} m_{\textup{Pl}}$ so that we have agreement with observations for the standard dispersion relation. Moreover, in this case the initial conditions at the bounce for the scalar field are: $\phi(0) = 1.033\,m_{\textup{Pl}}$ for model A \cite{AshtekarValues}, which produces again $141.30$ $e$-folds from the bounce until today; and $\phi(0) = 0.94\,m_{\textup{Pl}}$ for model B \cite{MRJ}, which now produces $130.19$ $e$-folds in total following a procedure entirely analogous to the Starobinsky case, see footnote~\ref{fn:calculations}. This being said,  the corresponding number of $e$-folds of the pre-inflationary and inflationary periods for the quadratic potentials are displayed in Table~\ref{Table. efolds Star} along with the Starobinsky potential analog values. As we can see, within  each model the pre-inflationary stage lasts almost the same, independently of the potential we are using, so that the concrete shape of the potential does not severely influence the pre-inflationary stage, as stated at the end of Sec.~\ref{LQC Background}. However, the potential does leave a trace in the inflationary period, changing the number of $e$-folds within this era. In particular, although inflation lasts more in model A than in model B for both potentials, Starobinsky potential produces more inflation in model A whereas quadratic potential does so in model B.

Focusing now on the analysis of perturbations, within model A the results we would obtain would be the same as with the Starobinsky potential, since we have the same number of $e$-folds from the bounce until today by definition and the background evolution is essentially the same (it only changes the duration of inflation, which now diminishes, but all the modes exit the horizon during this period). In the case of model B, for the quadratic potential we have $130.19$ $e$-folds, which differs from the $131.825$ of the Starobinsky potential. This means that the observable window with the quadratic potential in model B would be slightly more infrared than with the Starobinsky potential and, as a consequence, it would be less adiabatic but also further away from the ultraviolet scale, hence producing the same adiabaticity coefficient regardless of the dispersion relation and avoiding any kind of trans-Planckian problem while being compatible with observations.

\section{Conclusions}\label{Conclusions}

In this work, we have dealt with the trans-Planckian problem in models of LQC with cosmological perturbations. In~order to describe trans-Planckian effects, we have considered two different modified dispersion relations above some ultraviolet scale~$\kappa_c$, namely, the ones introduced by Unruh and by Corley and Jacobson. We have argued that the imprints that these modifications may leave in the power spectrum can be studied in terms of the adiabaticity coefficient of the modes while they were trans-Planckian. In particular, we have found that this parameter is of order $H / \kappa_c$, so that when this quantity is very small we do not expect any modification in the power spectrum.  This is the case during inflation, but not in the pre-inflationary dynamics usually obtained in LQC models.

This work complements the one carried out in~\cite{Francesesmdr}, where the power spectra with modified dispersion relations were computed in different LQC approaches to perturbations, but no insight was placed in understanding the origin of the modifications to those power spectra. 

We have investigated two concrete scenarios within hybrid LQC. The first one taking the initial conditions determined in~\cite{AshtekarValues} (although that work followed the dressed metric approach instead), which we named model A, and the second adopting the initial conditions of~\cite{MRJ}, which we called model B. These two models have been motivated in their own right within LQC and provide different primordial power spectra and $e$-folds of inflation. Given the total number of $e$-folds from the bounce until today, we first tracked the observable window in both models to determine if observable modes were ever trans-Planckian, particularly during the pre-inflationary epoch. Then we analysed the evolution of the adiabaticity coefficient of the most ultraviolet observable mode given the different dispersion relations.

This analysis shows that the existence of the trans-Planckian problem depends on the model we are using, both being compatible with observations for the linear dispersion relation. On the one side, model A faces a trans-Planckian problem, albeit arguably a soft one, as the adiabaticity coefficient becomes non-negligible for the Unruh dispersion relation when a large portion of the observable window is trans-Planckian. Moreover, the value of this coefficient for the trans-Planckian modes  strongly depends  on the dispersion relation adopted and so will the primordial power spectrum. By contrast, in model B we find that even though observable modes  lose adiabaticity during the pre-inflationary epoch, they are not trans-Planckian in that period, and so the primordial power spectrum is not sensitive to modifications of the dispersion relation. In this case there is no trans-Planckian issue, according to our analysis. This is fundamentally due to the fact that the initial conditions adopted in this model generate approximately $ 11\ e$-folds less of inflation. In general, with such a background it would be expected that the departure from near-scale-invariance in the primordial power spectrum already with the standard dispersion relation would spoil agreement with observations. However, the particular vacuum chosen in~\cite{MRJ} assures that the only departure with respect to near-scale-invariance is a power suppression of infrared modes, that seems to be favored by Planck data. Our analysis reveals that, in general, the trans-Planckian problem should not be disregarded in LQC as observable modes might be trans-Planckian in the pre-inflationary epoch and may lose the adiabaticity which is required for the trans-Planckian effects to be erased during inflation.

Nevertheless, this analysis presents some limitations that are important to mention. These are mainly three. The first one is that we have restricted ourselves to two concrete modified monotonic dispersion relations to account for trans-Planckian effects. These modifications were first introduced in the context of black holes ~\cite{Jacobson:1999zk, Unruh:1994je, Corley:1996ar} from a rather phenomenological perspective. It would be interesting to introduce trans-Planckian  modifications directly rooted at the quantum nature of geometry, following geometrical arguments of Loop Quantum Gravity.
In addition to this, we have chosen a specific ultraviolet scale, namely, the one that is related to the curvature radius at the bounce of LQC. This choice above others seems to be justified, but other choices could be made. 
Lastly, in this work we have only carried out a qualitative analysis of the trans-Planckian problem, for two particular models with specific numbers of $e$-folds and for a concrete potential, focusing on the evolution of the adiabaticity coefficient for trans-Planckian modes, rather than computing the power spectrum for completeness. We leave these calculations, as well as the consideration of other suitable potentials and models with different number of $e$-folds, for future research. However, the study carried out here anticipates what the results will~be, as we do not expect deep minima in the power spectrum at the initial stages of the expansion. Such a behavior could also lead to strong deviations in the primordial power spectrum, as shown in \cite{Fischer17} in an analogue condensed matter model to the de Sitter cosmos, given by an expanding quasi-two-dimensional Bose-Einstein condensate with dominant dipole-dipole interactions.

\acknowledgments

The authors would like to thank J. Olmedo for useful discussions.
This work has been supported by Project No. MICINN PID2020-118159GB-C44 from Spain.
R.N. acknowledges financial support in part from Fundação para a Ciência e a Tecnologia (FCT) through the research grant SFRH/BD/143525/2019 and from the Royal Society through the University Research Fellowship Renewal grant no. URF\textbackslash R\textbackslash221005.  M.L.G. acknowledges
financial support from IPARCOS through `Ayudas 2022 para la realización de Trabajos Fin de
Máster del Instituto de Física de Partículas y del Cosmos''.

\bibliography{BibTFM.bib}

\end{document}